\documentstyle[12pt]{article}
\begin{document} 
\global\parskip 6pt
\newcommand{\be}{\begin{equation}}
\newcommand{\ee}{\end{equation}}
\newcommand{\bea}{\begin{eqnarray}}
\newcommand{\eea}{\end{eqnarray}}
\newcommand{\non}{\nonumber}

\begin{titlepage}
\hfill{hep-th/0101194}
\vspace*{1cm}
\begin{center}
{\Large\bf Choptuik Scaling and Quasinormal Modes}\\
\vspace*{.2cm}
{\Large \bf in the AdS/CFT Correspondence}\\
\vspace*{2cm}
Danny Birmingham\footnote{E-mail: dannyb@pop3.ucd.ie}\\
\vspace*{.5cm}
{\em Department of Mathematical Physics\\
University College Dublin\\
Belfield, Dublin 4, Ireland}\\
\vspace{2cm}

\begin{abstract}
We establish an exact connection between the Choptuik scaling parameter
for the three-dimensional BTZ black hole, and the imaginary part of the
quasinormal frequencies for scalar perturbations. Via the AdS/CFT
correspondence, this leads to an interpretation of Choptuik scaling in
terms of the timescale for return to equilibrium of the dual conformal
field theory.
\end{abstract}
\vspace{1cm}
January 2001
\end{center}
\end{titlepage}

Within the context of numerical relativity, one of the most significant
recent results is the evidence for universal scaling behaviour in black
hole formation \cite{Chop, Gund}.
In particular, one considers a generic smooth one-parameter
family of initial data (labelled by $p$), such that a black
hole is formed for values of $p$ greater than a critical value $p^{*}$,
while no black hole is formed for $p< p^{*}$.
The mass of the black hole then satisfies
the scaling relation \cite{Chop}
\bea
M \sim (p - p^{*})^{\gamma},
\eea
where $\gamma$ is a universal exponent known as the
Choptuik scaling parameter.
In \cite{BS}, it was shown that
the Gott time machine \cite{Gott},
namely a two-body collision process, gives
a precise algebraic mechanism for the formation of the
$(2+1)$-dimensional
BTZ black hole.
This led to an exact analytic determination of
Choptuik scaling.

In \cite{Mann1}-\cite{wang3}, the quasinormal modes of scalar fields
in the background of anti-de Sitter black holes
were studied. The associated complex quasinormal frequencies
describe the decay of the scalar perturbation, and depend only on the
parameters of the black hole.
In terms of the $\mathrm{AdS/CFT}$ correspondence
\cite{Mald1}-\cite{Mald2}, an off-equilibrium configuration
in the bulk $\mathrm{AdS}$ space is related to an off-equilibrium
state in the boundary conformal field theory.
The timescale for the decay of the
scalar perturbation is given by the imaginary part of
the quasinormal frequencies.
Thus, by virtue of the $\mathrm{AdS/CFT}$ correspondence,
one obtains a prediction of the timescale for return to
equilibrium of the dual conformal field theory.
Interestingly, it was shown numerically \cite{Hor1} that the imaginary
part of the quasinormal frequencies for intermediate-sized
black holes, $\omega_{\mathrm{Im}}$,
scaled with the horizon radius, $r_{+}$.
In particular, it was found that
\bea
\omega_{\mathrm{Im}} \sim \frac{1}{\gamma} r_{+},
\label{hor}
\eea
where $\gamma$ is the Choptuik scaling parameter.
This relation, although not understood,
suggested a deeper connection between
black hole critical phenomena and quasinormal modes.

In this paper, we compute exactly the quasinormal modes
of massive scalar fields in the background of the BTZ
black hole, see also \cite{Gov, Lemos}.
It is shown that the imaginary part of the quasinormal
frequencies has a universal scaling behaviour precisely of the
form (\ref{hor}).
This leads to a conformal field theory interpretation
of Choptuik scaling within the context of
the $\mathrm{AdS/CFT}$ correspondence.

To begin, we recall that the line element for the BTZ black hole
can be written in the form \cite{BTZ1, Carlip}
\bea
ds^{2} = - \left(-M + \frac{r^{2}}{l^{2}} + \frac{J^{2}}{4r^{2}}\right)
dt^{2} + \left(-M + \frac{r^{2}}{l^{2}} + \frac{J^{2}}{4r^{2}}\right)^{-1}
dr^{2} + r^{2} \left(d\phi - \frac{J}{2r^{2}}dt\right)^{2}.
\label{metric}
\eea
The mass and angular momentum of the black hole can be expressed
in terms of the inner and outer horizon radii, $r_{\pm}$, as
\bea
M = \frac{r_{+}^{2} + r_{-}^{2}}{l^{2}},\;\;
J = \frac{2 r_{+}r_{-}}{l},
\eea
and we choose units for Newton's constant such that $8G = 1$.

We wish to study the properties of a massive scalar field in
the background geometry of the BTZ black hole. A special
feature of this $(2+1)$-dimensional case is that
the corresponding wave equation can be solved
exactly in terms of hypergeometric functions \cite{ghor,Satoh}.
By choosing appropriate boundary conditions,
we are led to an exact determination of the quasinormal modes
for the scalar field.
The scalar wave equation takes the form
\bea
\left(\nabla^{2} - \frac{\mu}{l^{2}}\right)\Phi = 0,
\eea
where $\mu$ is the mass parameter.
Using the ansatz,
\bea
\Phi = R(r) e^{-i\omega t} e^{im\phi},
\eea
with the change of variables
\bea
z = \frac{r^{2} - r^{2}_{+}}{r^{2} - r^{2}_{-}},
\eea
we are led to the radial equation
\bea
z(1-z)\frac{d^{2}R}{dz^{2}} + (1-z)\frac{dR}{dz} +
\left(\frac{A}{z} + B + \frac{C}{1-z}\right)R = 0.
\label{radial}
\eea
Here,
\bea
A &=& \frac{l^{4}}{4(r_{+}^{2} - r_{-}^{2})^{2}}(\omega r_{+}
- \frac{m}{l}r_{-})^{2},\non\\
B &=& - \frac{l^{4}}{4(r_{+}^{2} - r_{-}^{2})^{2}}(\omega r_{-}
- \frac{m}{l}r_{+})^{2},\non\\
C &=& -\frac{\mu}{4}.
\eea

We now define
\bea
R(z) = z^{\alpha}(1-z)^{\beta}F(z).
\eea
The radial equation then assumes the standard
hypergeometric form \cite{Ab}
\bea
z(1-z)\frac{d^{2}F}{dz^{2}} + [c - (1 + a + b)z]\frac{dF}{dz}
-abF = 0,
\label{hyp}
\eea
where
\bea
c &=& 2 \alpha + 1,\non\\
a+b &=& 2 \alpha + 2\beta,\non\\
ab &=& (\alpha
+ \beta)^{2} - B,
\label{coeff}
\eea
and
\bea
\alpha^{2} &=&-A,\non\\
\beta &=& \frac{1}{2}(1 \pm \sqrt{1 + \mu}).
\label{alpha}
\eea
Without loss of generality, we take
$\alpha = -i \sqrt{A}$ and $\beta = \frac{1}{2}
(1 - \sqrt{1 + \mu})$.

In the neighbourhood of the horizon, $z=0$, the two linearly
independent solutions of (\ref{hyp}) are given by \cite{Ab}
$F(a,b,c,z)$ and $z^{1-c}F(a-c+1,b-c+1, 2-c,z)$.
The quasinormal modes are defined
as solutions which are purely ingoing at the horizon,
and which vanish at infinity \cite{Hor1}.
The solution which has ingoing flux at the horizon is given by
\bea
R(z) = z^{\alpha}(1-z)^{\beta}F(a,b,c,z).
\eea
To implement the vanishing boundary condition at infinity, $z=1$,
we use the linear transformation formula \cite{Ab}
\bea
R(z) &=&
z^{\alpha}(1-z)^{\beta}(1-z)^{c-a-b}\frac{\Gamma(c)\Gamma(a+b-c)}
{\Gamma(a)\Gamma(b)}F(c-a,c- b,c-a-b+1,1-z) \non \\
&+& z^{\alpha}(1-z)^{\beta}\frac{\Gamma(c)\Gamma(c-a-b)}
{\Gamma(c-a)\Gamma(c-b)}F(a,b,a+b-c+1,1-z).
\label{infty}
\eea
Clearly, the first term in (\ref{infty}) vanishes. However,
the vanishing of the second term imposes the restriction
\bea
c-a = -n,\;\; {\mathrm or}\;\; c-b = -n,
\label{cond}
\eea
where $(n=0,1,2,...)$.
This condition leads directly to an exact
determination of the quasinormal
modes.
From (\ref{coeff}), we have
\bea
a &=& \alpha + \beta + i \sqrt{-B},\non\\
b &=& \alpha+\beta -i\sqrt{-B}.
\eea
Thus, we find that the left and right quasinormal modes, denoted by
$\omega_{L}$ and $\omega_{R}$, are given by
\bea
\omega_{L}
&=& \frac{m}{l} - 2i\;\left(\frac{r_{+}-r_{-}}{l^{2}}\right)
\left(n + \frac{1}{2}
+ \frac{1}{2}\sqrt{1 + \mu}\right),\non\\
\omega_{R}
&=& -\frac{m}{l} - 2i\;\left(\frac{r_{+}+r_{-}}{l^{2}}\right)
\left(n + \frac{1}{2}
+ \frac{1}{2}\sqrt{1 + \mu}\right).
\label{quasi}
\eea
It is important to stress that this is an exact calculation
of all quasinormal modes for the scalar field in a general
BTZ background. The result (\ref{quasi}) agrees with
the special cases considered in \cite{Gov, Lemos},
for $\mu = 0$ and $J=0$; quasinormal modes for the BTZ black
hole were first studied in \cite{Mann1}.
We also note that the imaginary parts of the quasinormal modes
scale linearly with the
left and right temperatures, defined by \cite{BSS}
$T_{L} = (r_{+} - r_{-})/2\pi l^{2}$
and $T_{R} = (r_{+} + r_{-})/2 \pi l^{2}$.

The aim now is to determine the precise connection between
these quasinormal modes and the Choptuik scaling parameter of the BTZ
black hole.
The first point to recall is that the BTZ black hole is defined
as a quotient of $\mathrm{AdS}_{3}$ by a discrete group of isometries
of $\mathrm{AdS}_{3}$ \cite{BHTZ}.
This is seen by noting \cite{Carlip}
that $\mathrm{AdS}_{3}$ can be viewed as the group manifold of
$SL(2,\mathbf{R})$, with isometry group  
$(SL(2,{\mathbf{R}}) \times SL(2,{\mathbf{R})})/Z_{2}$. 
Thus, for ${\mathbf{X}} \in SL(2,{\mathbf{R}})$, the isometry group
acts by left and right multiplication, ${\mathbf{X}} \rightarrow \rho_{L}
{\mathbf{X}}\rho_{R}$, with the identification  
$(\rho_{L},\rho_{R}) \sim (-\rho_{L}, -\rho_{R})$.
The BTZ black hole is then defined as the quotient
${\mathrm{AdS}}_{3}/\langle(\rho_{L},\rho_{R})\rangle$,
where the generators
$(\rho_{L},\rho_{R})$ are given by \cite{Carlip}
\bea
\rho_{L} = \left(\begin{array}{cc}
e^{\pi(r_{+} - r_{-})/l}&0\\
0&e^{-\pi(r_{+}-r_{-})/l}
\end{array}\right),\;\;
\rho_{R} = \left(\begin{array}{cc}
e^{\pi(r_{+} + r_{-})/l}&0\\
0&e^{-\pi(r_{+}+r_{-})/l}
\end{array}\right).
\eea

It is important to note
that elements of $SL(2, {\mathbf R})$ are classified according to
the value of their trace, namely \cite{Steif}
\bea
\mid{\mathrm{Tr}}\; T\mid &<& 2, \;\;
{\mathrm{Elliptic}},\non\\
\mid{\mathrm{Tr}}\; T\mid &=&2, \;\;
{\mathrm{Parabolic}},\non\\
\mid {\mathrm{Tr}} \;T\mid &>& 2, \;\; {\mathrm{Hyperbolic}}.
\eea 
Thus, we see that the generators of the BTZ black hole
are hyperbolic.

According to \cite{DJH1}, a point particle spacetime in
$(2+1)$ dimensions is defined
via identifications by an elliptic generator.
In \cite{Steif, Mats1, Mats2}, the formation of BTZ black holes
from point particle collisions was investigated.
In particular, it was shown \cite{BS}
that the Gott time machine \cite{Gott},
suitably generalized to anti-de Sitter space,
provides a precise mechanism
for the formation of the BTZ black hole. Moreover, this
purely algebraic process,
in which a product of two elliptic generators
becomes a hyperbolic generator, leads to an exact
analytic determination of the Choptuik scaling parameter.

The Gott time machine is defined as a two-body collision
process, with particles labelled by $A$ and $B$, such that the mass
and boost parameters obey a certain constraint, known as the Gott
condition.
The generator for each particle is defined in terms of
its mass and boost parameters, denoted by $\alpha$ and $\xi$.
Moreover, the effective
generator for the two particles
is given by the product \cite{DJH1, Carr, DJH2}, namely
$T^{G} = T_{B}T_{A}$.
The order parameter of interest is the trace of this generator,
which takes the form \cite{BS}
\bea
\frac{1}{2}\; {\mathrm{Tr}}\;T^{G} &=& -\cos \frac{\alpha_{A}}{2}\;\cos
\frac{\alpha_{B}}{2} - \sin \frac{\alpha_{A}}{2}\;\sin\frac{\alpha_{B}}{2}
\non\\
&+& \sin\frac{\alpha_{A}}{2}\;
\sin\frac{\alpha_{B}}{2}\left[ \cosh^{2}\left(\frac{\xi_{A} 
+ \xi_{B}}{2}\right) 
+\cosh^{2}\left(\frac{\xi_{A} - \xi_{B}}{2}\right)\right]\non\\
&-& \sin\frac{\alpha_{A}}{2}\;
\sin\frac{\alpha_{B}}{2}\cos(\phi_{A} - \phi_{B})
\left[ \cosh^{2}\left(\frac{\xi_{A} + \xi_{B}}{2}\right) 
-\cosh^{2}\left(\frac{\xi_{A} - \xi_{B}}{2}\right)\right].
\label{Gott}
\eea
The original Gott time machine is recovered by choosing particles
with equal masses, and equal and opposite boosts, namely
$\alpha_{A} = \alpha_{B} = \alpha, \xi_{A} = \xi_{B} = \xi, 
\phi_{A} - \phi_{B} = \pi$.
Thus, when the Gott condition is satisfied, namely
$\sin^{2}\frac{\alpha}{2}\;\cosh^{2} \xi > 1$, we see that
$T^{G}$ is a hyperbolic generator. When the Gott condition is
not satisfied, we have
an elliptic generator.

To construct the BTZ black hole,
we simply take the
independent left and right generators $\rho_{L},\rho_{R}$
to be defined in terms of two-particle Gott generators.
Thus, we take $\rho_{L} = T^{G}$ in (\ref{Gott}) with
$\alpha_{A} = \alpha_{B} = \alpha, \phi_{A} - \phi_{B} = 0$.
This gives
\bea
\frac{1}{2} \;{\mathrm{Tr}}\;\rho_{L} = \cosh\left
(\frac{\pi}{l}(r_{+} - r_{-}) \right) = -1 + 2 \sin^{2}\frac{\alpha}{2}\;
\cosh^{2}\left(\frac{\xi_{A} - \xi_{B}}{2}\right) \equiv p_{L}.
\label{chopl}
\eea
For the right generator, we choose $\rho_{R} = T^{G}$ with
$\alpha_{A} = \alpha_{B} = \alpha, \phi_{A} - \phi_{B} = \pi$,
leading to
\bea
\frac{1}{2} \;{\mathrm{Tr}}\;\rho_{R} = \cosh \left( \frac{\pi}{l}
(r_{+} + r_{-})\right)
= -1 + 2 \sin^{2}\frac{\alpha}{2}\;
\cosh^{2}\left(\frac{\xi_{A} + \xi_{B}}{2}\right) \equiv p_{R}.
\label{chopr}
\eea
We see that both $\rho_{L}$ and $\rho_{R}$
become hyperbolic if the input parameters $\alpha, \xi_{A}, \xi_{B}$
satisfy the appropriate Gott conditions, namely $p_{L}>1$ and
$p_{R} > 1$. Thus, the critical value of the input parameters
is $p_{L}^{*} = p_{R}^{*} = 1$.
As shown in \cite{BS}, the Choptuik scaling parameter can now
be simply read off from (\ref{chopl}) and (\ref{chopr}),
by using the formula,
${\mathrm{arccosh}}\; p = \ln [p + \sqrt{p^{2} - 1}]$.
Writing $p_{L} = p_{L}^{*} + \epsilon$,
and $p_{R} = p_{R}^{*} + \epsilon$,
we find to leading order
\bea
\frac{r_{+} - r_{-}}{l} &=& \frac{\sqrt{2}}{\pi} (p_{L} -
p_{L}^{*})^{1/2},\non\\
\frac{r_{+} + r_{-}}{l} &=& \frac{\sqrt{2}}{\pi}
(p_{R} - p_{R}^{*})^{1/2}.
\eea
Thus, the Choptuik scaling parameter for $(r_{+}\pm r_{-})$ is
$\gamma = 1/2$. A scaling exponent of $1/2$ was also found
for collapsing dust shells in \cite{Peleg}. Other aspects of
Choptuik scaling for the BTZ black hole have been investigated in
\cite{Pret}-\cite{clement}.

We can now compare this result with the quasinormal
frequencies (\ref{quasi}). We see immediately that
the negative of the imaginary part of $\omega_{L}$
and $\omega_{R}$, denoted by $(\omega_{L})_{\mathrm{Im}}$
and $(\omega_{R})_{\mathrm{Im}}$, scales with $(r_{+} - r_{-})$
and $(r_{+} + r_{-})$, respectively. In particular, we have
\bea
(\omega_{L})_{\mathrm{Im}} &=& \frac{1}{\gamma}
\left(\frac{r_{+} - r_{-}}{l^{2}}\right)
\left(n + \frac{1}{2} + \frac{1}{2}\sqrt{\mu + 1}\right),\non\\
(\omega_{R})_{\mathrm{Im}} &=& \frac{1}{\gamma}
\left(\frac{r_{+} + r_{-}}{l^{2}}\right)
\left(n + \frac{1}{2} + \frac{1}{2}\sqrt{\mu + 1}\right).
\eea

We have thus established an exact connection between the Choptuik
scaling parameter and the imaginary part of the quasinormal modes.
It is satisfying that in the $(2+1)$-dimensional case, these exact
calculations lead to a result precisely of the form
noticed in \cite{Hor1}.
By virtue of the $\mathrm{AdS/CFT}$ correspondence, the imaginary part
of the quasinormal modes has a direct interpretation in
the dual conformal field theory. In the case at hand, the boundary
conformal field theory of $\mathrm{AdS}_{3}$ contains both left-moving
and right-moving sectors \cite{BH}, with Virasoro generators
$\bar{L}_{0} = (r_{+}-r_{-})^{2}/2l$ and
$L_{0} = (r_{+} + r_{-})^{2}/2l$, respectively.
Thus, the return to equilibrium of the conformal field theory is
specified in terms of the left and right timescales given by
$\tau_{L} = 1/(\omega_{L})_{\mathrm{Im}}$ and
$\tau_{R} = 1/(\omega_{R})_{\mathrm{Im}}$.
Further analysis of BTZ black hole formation within the context
of the $\mathrm{AdS/CFT}$ correspondence has been presented
in \cite{Bal}-\cite{Louko}.

\end{document}